\documentclass[12pt,pra,aps]{revtex4}

\usepackage[english]{babel}
\usepackage{latexsym}
\usepackage{graphicx}
\usepackage{subfigure}
\usepackage{epsfig}
\usepackage{amsfonts}
\usepackage{amssymb}
\usepackage{amsmath}
\usepackage{bm}

\begin{document}
\title{Integrability vs Quantum Thermalization}
\author{Jonas Larson}
\address{Department of Physics,
Stockholm University, AlbaNova University Center, Se-106 91 Stockholm,
Sweden}
\email{jolarson@fysik.su.se}
\begin{abstract}
Non-integrability is often taken as a prerequisite for quantum thermalization. Still, a generally accepted definition of quantum integrability is lacking. With the basis in the driven Rabi model we discuss this careless usage of the term ``integrability'' in connection to quantum thermalization. The model would be classified as non-integrable according to the most commonly used definitions, for example, the only preserved quantity is the total energy. Despite this fact, a thorough analysis conjectures that the system will not thermalize. Thus, our findings suggest first of all ($i$) that care should be paid when linking non-integrability with thermalization, and secondly ($ii$) that the standardly used definitions for quantum integrability are unsatisfactory.   
\end{abstract}
\maketitle

\section{Introduction}
The concept of integrability is well defined in classical systems~\cite{arnald,gutzwiller}. Integrability here means that the number of degrees of freedom is smaller than the number of {\it independent} constants of motion. Constants of motions in classical systems are characterized by vanishing {\it Poisson brackets}, and independence by mutually vanishing Poisson brackets. Classical integrability implies that the solutions are periodic and live on a torus of constant energy~\cite{com1} in phase space. Translating the above definition to quantum Hamiltonian systems directly leads to complications and there is no accepted definition of integrability in quantum systems~\cite{qint}. As an example, the number of degrees of freedom for quantum systems is generally taken as the dimension of the Hilbert space, and in particular the Hilbert space dimension can be finite. Classically, the degrees of freedom are necessarily continuous variables and there seem to be a contradiction in having a well defined quantum-classical correspondence, i.e. finite size Hilbert space systems do not have a proper classical limit. The spin, for example, is a pure quantum property.  

With the development of techniques in isolating and controlling quantum systems~\cite{maciek,qcon,pol}, questions regarding quantum integrability have gained renewed interest. Of special interest is out of equilibrium dynamics in closed quantum systems~\cite{pol}. Cold atom systems are especially practical for {\it in situ} measurements of quantum many-body systems and they thereby also provide a handle to study pure quantum evolution~\cite{exp1}. As a result, long standing questions in quantum statistical mechanics can now be addressed in an experimentally controlled way. Of particular interest is the long time evolution and whether an interacting quantum system equilibrates and if so what characterizes the relaxed state~\cite{eth}. A common believe is that for a non-integrable quantum system the state {\it thermalizes}, by which we mean that expectations of any local observable $\hat{A}$ can be evaluated from a micro-canonical state $\hat{\rho}_\mathrm{MC}$. This conjecture has been supported in several numerical studies of various models~\cite{gge,thermo1}. However, it seems specious to coin such an assumption based on a concept that still today lacks a proper definition. Moreover, it has been numerically demonstrated that using standard definitions for quantum integrability one can find models that are non-integrable and still do not thermalize~\cite{eisert1}. The present work adds to this reference.

Quantum thermalization has become deeply connected to interacting many-body systems~\cite{pol}. It is important to understand, however, that there is nothing in the theory that relies on having a quantum many-body system, i.e. a system possessing many degrees of freedom. It is rather properties of the eigenstates and the spectrum that determine the fate of the state~\cite{eth}. Indeed, quantum thermalization has been demonstrated in systems whose classical counterparts possess only two degrees of freedom~\cite{altland}. In the works of ref.~\cite{altland}, a common feature is instead that the corresponding classical models are chaotic~\cite{haake}. A question thereby rises: Is classical chaos a common feature of systems that quantum thermalize? Naturally, this cannot be a general condition since, as argued above, some quantum systems do not have a well defined classical limit. For example, the work~\cite{eisert1} considers a disorder Heisenberg spin-1/2 chain for which a proper classical limit does not exist. To spur the discussion about quantum integrability and thermalization, in this work we consider a purely quantum model (i.e. lacking a classical counterpart) that does not obey standard criteria for integrability and still do not show any signatures of thermalization. More precisely, we analyze out of equilibrium long term evolution in the {\it driven Rabi model} (RM) which describes interaction between a spin-1/2 system and a boson mode. After a general discussion about integrability and thermalization we investigate statistical properties of various local expectation values.

\section{Integrability and Quantum Thermalization}
\subsection{Quantum Integrability}
Already mentioned in the introduction, integrability in classical systems has a clear meaning. An $N$-dimensional Hamiltonian system $H({\bf p},{\bf q})$ is said to be integrable if: ($i$) there exist $N$ single-valued constants of motion $I_n$, i.e. $\{I_n,H\}=0$, where $\{\,\,,\,\}$ denotes the Poisson bracket, ($ii$) the constants of motions $I_n$ are functionally independent, and ($iii$) the constants of motion $I_n$ are {\it in involution} meaning $\{I_n,I_{n'}\}=0$, $\forall\,\, n,n'$. For an integrable system, the solutions $({\bf p}(t),{\bf q}(t))$ are periodic and evolve on $(N-1)$-dimensional tori in phase space. For such constrained evolution, the solutions do only explore a small part of the phase space. When the integrability condition is (slightly) lifted, the tori start to deform in accordance with {\it KAM-theory} (Kolmogorov-Arnold-Moser)~\cite{gutzwiller}. The solutions are (in general~\cite{com}) no longer periodic and cover a larger part of phase space. This describes the transition from regular to chaotic motion in classical systems.      

Trying to define integrability for a Hamiltonian quantum systems $\hat{H}$ is far from trivial~\cite{qint,qint2,qint3}. There have been numerous different attempts to give a meaningful and consistent definition. We summarize some of the more traditional ones in the following list.

\begin{enumerate}
\item {\it Traditional I.} The far most commonly used definition for quantum integrability is obtained from translating the classical definition into a quantum language. Thus, functions $I_n$ are replaced by operators $\hat{I}_n$ and Poisson brackets by commutators $\{\,\,,\,\}\rightarrow i[\,\,,\,]/\hbar$. It is easy to reject such a definition by noticing that the projectors $\hat{P}_\alpha=|\psi_\alpha\rangle\langle\psi_\alpha|$, with $|\psi_\alpha\rangle$ a (non-degenerate) eigenstate of the Hamiltonian define constants of motion and mutually commute. Thus, it is possible to find a set of constants of motion such that any quantum system appears integrable. In addition, it can be proven that for a set $\{\hat{I}_n\}$ of commuting operators there exist a single operator $\hat{I}$ and a set of functions $f_n(x)$ such that $\hat{I}_n=f_n(\hat{I})$~\cite{vN}. This theorem tells us that one should specify what is ment by ``the number of independent operators''.  

\item {\it Traditional II.} The problem with the above definition led to the notion of {\it relevant} and {\it irrelevant} constants of motion~\cite{qint2,qint4}. The relevant constants of motion are those which can be associated with a classical counterpart. This again have some flaws since inequivalent quantum constants of motion can share the same classical limit~\cite{qint2}, and not all quantum systems do have a classical limit to start with. 

\item {\it Scattering.} A quantum system is integrable if its scattering is non-diffractive~\cite{qint3}. This applies only to continuous models and it relies on properties of the asymptotic scattered states. Thinking in terms of a scattering problem, if the outgoing solution contains ``diffractive contributions'' the system is non-integrable. 

\item {\it Bethe solution.} A quantum system is integrable if it can be solved with the {\it Bethe ansatz}. A definition of this type cannot be general since there exists models that are solvable without a Bethe ansatz. Furthermore, as for the previous definition the present one originates from systems of many interacting particles. We wish to have a general definition that is independent on the particle number or the number of degrees of freedom. 

\item {\it Poissonian level statistics.} A quantum system is integrable if its energy level statistics is Poissonian~\cite{berry}. Following ref.~\cite{berry}, this definition relies on semi-classical arguments and to systems with continuous degrees of freedom. Thus, it is not general for any systems.

\item {\it Level crossings.} A quantum system is integrable if it shows level crossings. This definition is related to the previous one since avoided crossings are characteristic for systems showing {\it level repulsion}, i.e. the energy level statistics follows a {\it Wigner-Dyson distribution}~\cite{haake}. Note that the definition does not say anything about avoided crossings.

\item {\it Solvability.} A quantum system is integrable if it is exactly solvable. 

\end{enumerate}

It can be argued that the defining properties of (iii), (v) and (vi) are rather consequences of non-integrability than defining it. The usefulness of definitions (iv) and (vii) may be discussed (for obvious reasons). We should mention that the list above is not complete, there exist further definitions not included here~\cite{qi,braak}. 

\subsection{Quantum Thermalization}
In recent years we have seen an increased interest in dynamics of closed quantum systems~\cite{pol}. An open question with a very long history concerns equilibration of such states~\cite{sch}. A central topic in this field has been to understand local relaxation to a thermal state of a quantum many-body state~\cite{eth,gge,thermo1,eisert1,thermo2}. To gain deeper insight in the mechanism driving quantum thermalization, several concepts have been introduced, for example: the {\it eigenstate thermalization hypothesis} (ETH)~\cite{eth}, quantum central limit theorems~\cite{eisert2}, system-bath entanglement~\cite{ent} and the {\it eigenstate randomization hypothesis}~\cite{erh}. 

Especially the ETH has been thoroughly studied. To explain the idea of ETH, let us express the time evolution of some state $|\psi(t)\rangle$ in terms of eigenstates $|\psi_\alpha\rangle$ of the Hamiltonian,
\begin{equation}\label{expand}
|\Psi(t)\rangle=\sum_\nu C_\nu e^{-iE_\alpha t/\hbar}|\psi_\nu\rangle.
\end{equation}
The expectation of some observable $\hat{A}$ reads
\begin{equation}
\langle\hat{A}(t)\rangle=\sum_\nu|C_\alpha|^2A_{\nu\nu}+\sum_{\nu\neq\mu}C_\nu^*C_\mu e^{i(E_\nu-E_\mu)t/\hbar}A_{\nu\mu},
\end{equation}
where $A_{\nu\mu}=\langle\psi_\nu|\hat{A}|\psi_\mu\rangle$. If the state equilibrates, the long time expectation $\langle\hat{A}\rangle^\mathrm{LT}$ should attain the time averaged value
\begin{equation}
\langle\hat{A}\rangle^{\mathrm{LT}}=\lim_{T\rightarrow \infty}\frac{1}{T}\int_0^Tdt\langle\hat{A}(t)\rangle=\sum_\nu|C_\nu|^2A_{\nu\nu}.
\end{equation}
This expectation is obtained when the long time state is diagonal in the eigenvalue basis $\hat{\rho}_\mathrm{LT}=\sum_\nu|C_\nu|^2|\psi_\nu\rangle\langle\psi_\nu|$. For the situations of interest for us, the probabilities $|C_\nu|^2$ are only non-zero in some energy window $\Delta E$ around $E=\langle\Psi|\hat{H}|\Psi\rangle$. The number of populated states $|\psi_\nu\rangle$ in the sum (\ref{expand}) can be estimated with the {\it inverse participation ratio}~\cite{ipr}
\begin{equation}\label{partrat}
\eta_\psi=\left(\sum_\nu|C_\nu|^4\right)^{-1}.
\end{equation} 
Clearly, $\eta_\psi\gg1$ in order to expect equilibration. For a microcanonical distribution, $\hat{\rho}_\mathrm{MC}=N(E,\delta)^{-1}\sum_{\gamma\in\delta}|\psi_\gamma\rangle\langle\psi_\gamma|$ where $\delta$ ($>\Delta E$) is again an energy window around $E$ and $N(E,\delta)$ being the number of states within $\delta$, the expectation become
\begin{equation}
\langle\hat{A}\rangle^{MC}=\frac{1}{N(E,\delta)}\sum_{\gamma\in \delta}A_{\gamma\gamma}.
\end{equation}
The state thermalizes if $\langle\hat{A}\rangle^\mathrm{LT}=\langle\hat{A}\rangle^\mathrm{MC}$ up to corrections of the order $\mathcal{O}(\eta_\psi^{-1})$. Now, the ETH says that for a state that thermalizes, $A_{\alpha\alpha}$ varies little within the energy interval $\Delta E$. We directly see that if $A_{\alpha\alpha}$ is more or less constant in the interval of interest, the expectation $\langle\hat{A}\rangle^\mathrm{LT}$ approximates $\langle\hat{A}\rangle^\mathrm{MC}$ which you obtain from the microcanonical distribution. Thus, ETH predicts that for a system that supports thermalization (given the initial energy), any functions $\langle\hat{A}\rangle$ has a weak $E$-dependence on the scale $\Delta E$.

The ETH does not say whether a system will thermalize or not, it is rather a property of a system that thermalizes~\cite{com2}. Without deeper reflection, it is often assumed that any non-integrable system will thermalize. From the discussion in the previous subsection it is clear that there is a great ambiguity in such an assumption, simply because there is no generally accepted definition of quantum integrability. The problem might be circumvented for systems with a well defined classical limit, and it has indeed been found that several systems were the corresponding classical counterparts are chaotic do thermalize. While numerical experience indicates such a fact, there is no strict proof that this is true in general. The situation is more complicated when the system of interest does not allow for a simple classical limit. Note that here chaos is discussed in terms of the classical model, i.e. chaos defined from a positive Lyapunov exponent. The connection between quantum chaos, defined from level statistics of the energy spectrum, and quantum thermalization has been discussed~\cite{qcth1,qcth2}. It was particularly found that whether the state will thermalize or not depends strongly on the initial energy~\cite{qcth2}. For example, if the state populates predominantly eigenstates corresponding to energies at the edges of the spectrum thermalization is typically absent.   

Of course, the discussion above on the ETH is fully general, i.e. there are no assumptions on number of degrees of freedom nor on existence of a classical limit. As the name suggests, it relies on the properties of the eigenvectors. In the next section we will study a particular model which should not be classified as integrable according to the definitions above, and still we find no indications of quantum thermalization.   

\section{Generalized Rabi model}
\subsection{Driven Rabi Model}
The RM~\cite{rabi} has a long history in quantum optics and especially in {\it cavity quantum electrodynamics} (QED)~\cite{jonas0}. Despite its simplicity, a spin-$1/2$ system coupled to a single boson mode, the physics is extremely rich. In most experiments to date, both in cavity and circuit QED, the {\it rotating wave approximation} (RWA) is well justified and the RM is then approximated with the exactly solvable {\it Jaynes-Cummings model}~\cite{knight}. Within the RWA, the number of excitations $\hat{N}=\hat{a}^\dagger\hat{a}+\hat{\sigma}_z/2$ ($\hat{a}^\dagger$ and $\hat{a}$ are the creation and annihilation operators for the boson mode and $\hat{\sigma}_z$ is the Pauli $z$-matrix acting on the spin) is preserved which implies a continuous $U(1)$ symmetry, i.e. $[\hat{U}_\phi,\hat{H}_{JC}]=0$ with $\hat{U}_\phi=e^{i\hat{N}\phi}$. More recently, an alternative RWA was considered in order to derive an analytically solvable model with a larger validity regime compared to the regular RWA~\cite{irish}. Also this time, the applied approximation results in restoring the same $U(1)$ symmetry. Relaxing the RWA means that the $U(1)$ symmetry is broken down to a discrete $Z_2$ parity symmetry ($[\hat{U}_{\pi},\hat{H}_R]=0$)~\cite{parity}. In the spirit of the previous section, it is not clear whether a discrete symmetry should imply integrability of the RM. Furthermore, while the boson mode has a well defined classical limit the spin does not and one cannot thereby define integrability from any classical limit.

The search for a solution of the RM has a long history~\cite{rabisol1}. A breakthrough came in 2011 when D. Braak claimed to have solved the RM~\cite{braak}. In particular, the spectrum can be divided into a {\it regular} and an {\it exceptional part}. The regular part is given by zeros of a transcendental function. The exceptional solutions have a simple analytical expression but, on the other hand, they only exist for certain system parameters. More recently, A. Moroz remarked that the RM is not exactly solvable~\cite{moroz}, but rather an example of a quantum model that is {\it quasi-exactly solvable}~\cite{quasi}. Thus there is a debate whether the RM is in fact solvable or not.

We may break the $Z_2$ symmetry of the RM by considering an external driven,
\begin{equation}\label{drabi}
\hat{H}=\hat{a}^\dagger\hat{a}+\frac{\omega}{2}\hat{\sigma}_z+g\left(\hat{a}^\dagger+\hat{a}\right)\hat{\sigma}_x+\lambda\hat{\sigma}_x.
\end{equation}
Here, we have introduced dimensionless parameters by letting the energy $\hbar\Omega$ of a single boson set a characteristic energy scale, $\omega$ is the energy separation of the two spin states $|1\rangle$ and $|2\rangle$, $\hat{\sigma}_x$ is the Pauli $x$-matrix, $g$ is the spin-boson coupling, and finally $\lambda$ is the drive amplitude. By letting $\lambda=0$ we regain the Rabi Hamiltonian $\hat{H}_R$. The drive term breaks the parity symmetry since $\hat{U}_\pi\hat{\sigma}_x\hat{U}_\pi^{-1}=-\hat{\sigma}_x$. Note that the drive of the spin can be removed by unitarily transform the Hamiltonian with the displacement $\hat{U}=e^{-\frac{\lambda}{\sqrt{2}g}\left(\hat{a}^\dagger-\hat{a}\right)}$. In return, the transformed Hamiltonian contains a drive of the boson mode, i.e. $\left(\hat{a}^\dagger+\hat{a}\right)\lambda/g$. Thus, driving of the spin or the boson mode is unitarily equivalent and here we consider the first option. Judging from refs.~\cite{braak} and~\cite{quasi}, it seems that also the driven Rabi model (\ref{drabi}) is of the quasi-exactly solvable type. This fact may naturally be of importance in terms of thermalization.

Let us return to the definition of quantum integrability in the previous subsection and check whether the driven Rabi model fulfills any of them. 

\begin{figure}[h]
\centerline{\includegraphics[width=10cm]{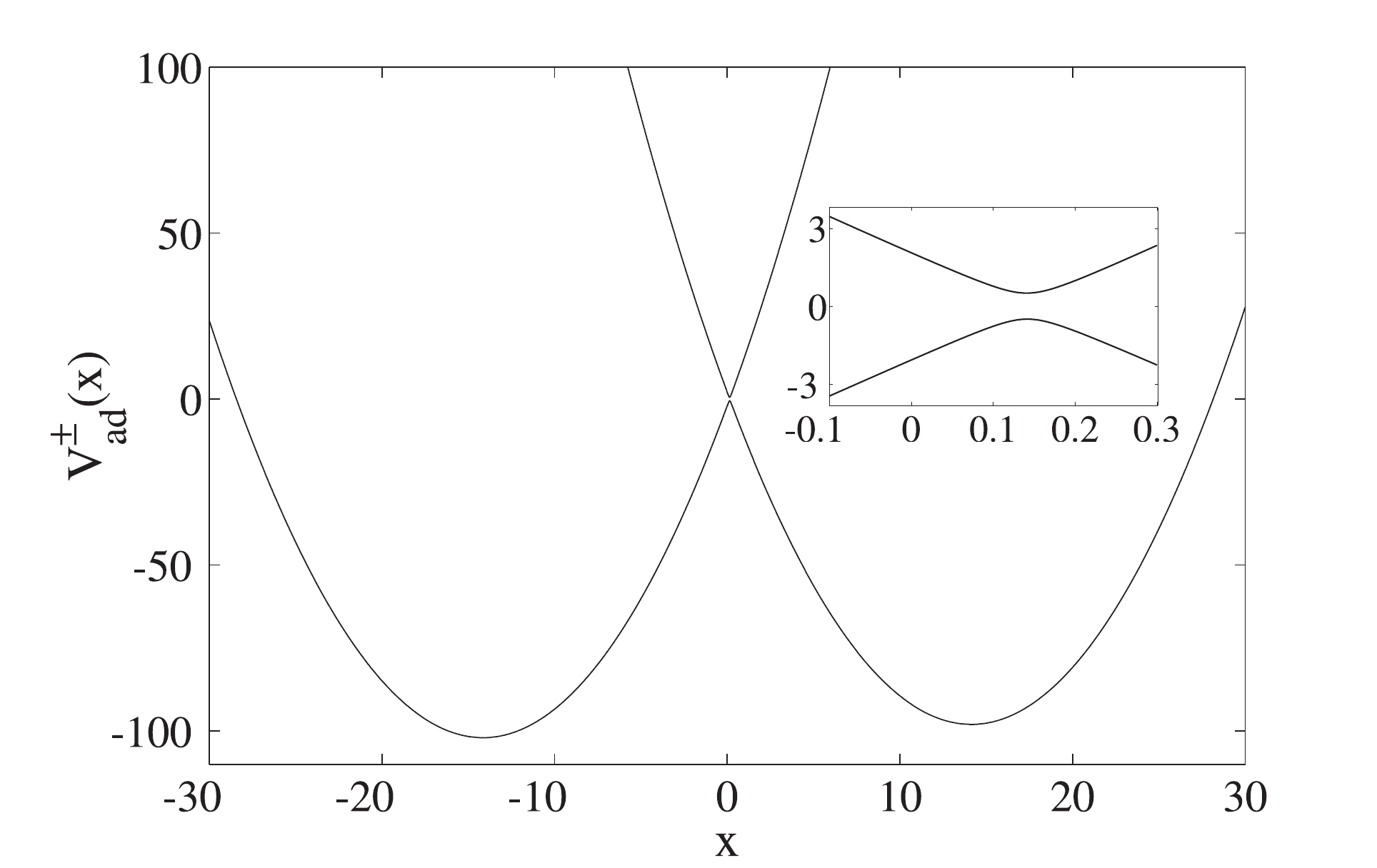}}
\caption{The two adiabatic potentials $V_\mathrm{ad}^\pm(x)$ for the parameters $\omega=1$, $g=10$ and $\lambda=2$. The inset shows a zoom of the avoided crossing.} \label{fig1}
\end{figure}

\begin{enumerate}

\item {\it Traditional I.} As already pointed out, this definition is pointless since one can always find a set of constants of motion such that any system would be considered integrable.

\item {\it Traditional II.} With the driving, the $Z_2$ symmetry is broken and the only relevant constant of motion is the energy. In this respect, the driven Rabi model should not be classified as integrable. Of course, we have a problem here since the spin does not have a natural classical limit. We may, however, perform a semi-classical approximation in which the boson mode is treated at a mean-field level, while the spin is still kept as a quantum entity. Thus, we make a coherent state ansatz for the boson field where the bosonic operators are replaced with their corresponding coherent amplitudes, $\hat{a}\rightarrow\alpha$ and $\hat{a}^\dagger\rightarrow\alpha^*$. In doing this we neglect any quantum correlations between the spin and the boson mode. As a result~\cite{com4}, a generic spin state can be written $|\Theta\rangle=\left[\sqrt{\frac{1+Z}{2}}\,,\,\,\sqrt{\frac{1-Z}{2}}e^{i\Delta_\phi}\right]^T$, where $Z$ is the {\it inversion} ($\langle\hat{\sigma}_z\rangle=Z$) and $\Delta_\phi$ the {\it relative phase} ($\tan(\Delta_\phi)=\langle\hat{\sigma}_y\rangle/\langle\hat{\sigma}_x\rangle$). By introducing quadratures $x$ and $p$ according to $\alpha^*=(x+ip)/\sqrt{2}$ and $\alpha=(x-ip)/\sqrt{2}$, we can write a ``classical'' Hamiltonian
\begin{equation}\label{cham}
H_{cl}=\frac{p^2}{2}+\frac{x^2}{2}+\frac{\omega}{2}Z+\left(gx\sqrt{2}+\lambda\right)\sqrt{1-Z^2}\cos(\Delta_\phi).
\end{equation}
The semi-classical equations of motion now become      
\begin{equation}\label{ceom}
\begin{array}{llll}
\dot{x}=p, & & \dot{p}=-x-g\sqrt{2}\sqrt{1-Z^2}\cos(\Delta_\phi),\\ \\
\dot{Z}=gx\sqrt{2}\sqrt{1-Z^2}\sin(\Delta_\phi), & & \displaystyle{\dot{\Delta_\phi}=\frac{\omega}{2}-\left(g\sqrt{2}x+\lambda\right)\cos(\Delta_\phi)\frac{Z}{\sqrt{1-Z^2}}}.
\end{array}
\end{equation}
Putting $\lambda=0$ we obtain the classical equations of motion for the Dicke model which have been demonstrated to be chaotic~\cite{altland,dicke}. The corresponding semi-classical equations of motion for the RM were also analyzed in ref.~\cite{nagel} with clear signatures of chaos. See also ref.~\cite{jonas2} which studies similar semi-classical equations of motion. We have solved the equations of motion (\ref{ceom}) numerically and studied different Poincar\'e sections~\cite{strogatz}. For large enough couplings $g$ they all show well developed chaos. In this respect, the RM should not be considered integrable.

\begin{figure}[h]
\centerline{\includegraphics[width=10cm]{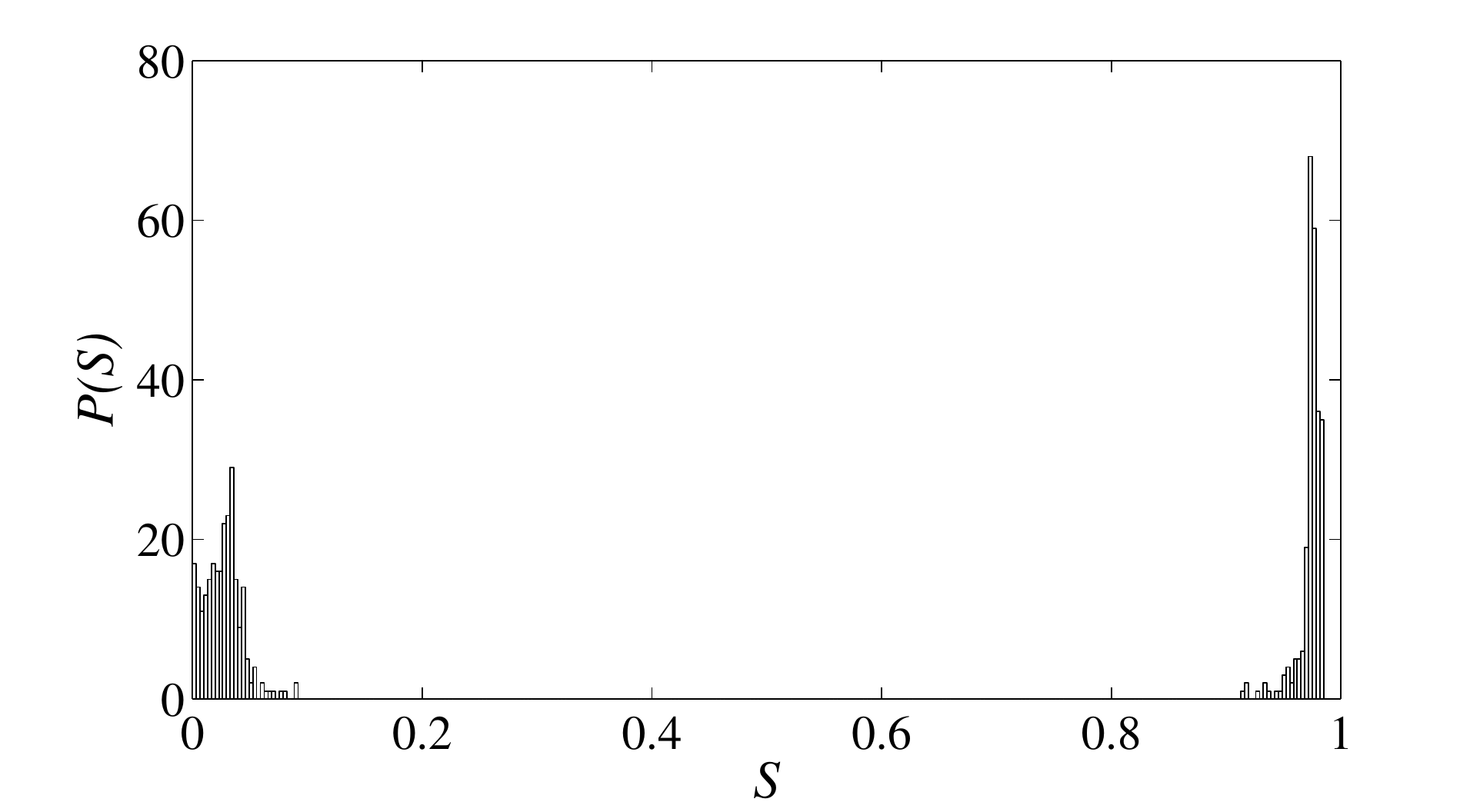}}
\caption{Level statistics of the driven Rabi model for the dimensionless parameters $\omega=1$, $g=10$ and $\lambda=2$. Energies $0<E<250$ have been considered.} \label{fig2}
\end{figure}

\item {\it Scattering.} Since the spectrum of $\hat{H}$ is discrete, the idea of non-diffractive scattering does not apply to our system.

\item {\it Bethe solution.} The Bethe ansatz is typically applied to quantum many-body problems with continuous degrees of freedom. Hence, we cannot apply such approaches to the RM.

\item {\it Poissonian level statistics.} Level statistics explores the distribution $P(S)$ - the number of energies with certain nearby energy gaps $S_n=E_{n+1}-E_n$. Typical for systems showing regular dynamics is that the level statistic of the spectrum follows a Poisson distribution $P(S)=e^{-S}$. Characteristic for chaotic systems, on the other hand, is the level-repulsion effect and statistics is normally given by a Wigner-Dyson distribution $P(S)=(S\pi/2)e^{-S^2\pi/4}$~\cite{haake}. Indeed, the level repulsion is often used as a definition for quantum chaos~\cite{haake,dicke}. The level statistics of the RM has been studied in the past~\cite{kus}. Despite the similarity to the Dicke model, their statistics are very different. While the Dicke model shows clear level-repulsion in the chaotic regime~\cite{dicke}, the level statistics of the RM is neither of Poisson nor Wigner-Dyson shape. This was also pointed out by D. Braak in \cite{braak} where he noticed that the energies are rather equally spaced throughout. Many of the properties of the spectrum can be understood within the {\it Born-Oppenheimer approximation} (BOA)~\cite{jonas0,BO,jonas3}. In the BOA we decouple the internal degrees of freedom of $\hat{H}$ by diagonalizing the spin part of eq.~(\ref{drabi}). The two resulting {\it adiabatic potential curves} for the driven RM become~\cite{jonas0,jonas3}
\begin{equation}
V_\mathrm{ad}^\pm(x)=\frac{x^2}{2}\pm\sqrt{\frac{\omega^2}{4}+\left(\sqrt{2}gx+\lambda\right)^2}.
\end{equation} 
The two potentials are displayed in fig.~\ref{fig1}. We see that in this {\it ultrastrong coupling regime} ($g>\sqrt{\omega}$), the lower adiabatic potential $V_\mathrm{ad}^-(x)$ has a double-well structure. This symmetric structure reflects the $Z_2$ parity symmetry, which implies that for the double-well potential the spin states are ``opposite'' between the two potential wells. The driving causes the double-well to be asymmetric, and hence the $Z_2$ symmetry is broken. The $\hat{\sigma}_z$ term in the Hamiltonian opens up a gap between the two potentials (see the inset of the figure). Around this avoided crossing, the BOA is likely to break down and it is no longer possible to think about the system as two decoupled potentials. For $\lambda=0$, the double-well potential is symmetric and for large couplings $g$ the spectrum is to a good approximation degenerate for energies $E<0$. For positive and moderate energies, this quasi degeneracy is lost. These are properties also shared by the Dicke model and there the double-well structure characterizes the Dicke phase transition and the corresponding spontaneous breaking of the $Z_2$-symmetry~\cite{brenacke}. For even larger energies, the anhorminicity deriving from the spin-boson coupling becomes extremely weak and the two potentials are approximately harmonic. For a large driving, i.e. $\lambda>g$, the asymmetry of the double-well potential is distinct, which will split the quasi degeneracy. Nevertheless, provided that $g$ is large the negative energies can be approximated with those of two harmonic oscillators. Taking all these aspects into account, we draw the conclusion that in order to find any non-trivial level statistics the spectrum should be explored for moderate and positive energies. This has also been confirmed numerically, i.e. the largest deviation from Poissonian statistics is regained in this energy regime. In fig.~\ref{fig2} we show the distribution $P(S)$ of the driven RM for energies $0<E<250$ and for the same parameters as in fig.~\ref{fig1}. The pronounced ``clustering'' clearly demonstrate the absence of Poissonian statistics. The clustering at small $S$ is even indicating some level repulsion. 

As a remark on level statistics. It can be shown that the RM is deeply connected to the $E\times\varepsilon$ Jahn-Teller model~\cite{jt}. While the $E\times(\beta_1+\beta_2)$ model shows full blown quantum chaos~\cite{eva1}, the $E\times\varepsilon$ model displays classical chaos and some `incipience' of quantum chaos~\cite{eva2}. 

\item {\it Level crossings.} Parameter dependence of the spectrum of the RM was studied in~\cite{cross}. In contrast to the solvable Jaynes-Cummings model~\cite{knight}, the energies of the RM show typically avoided crossings within the two parity sectors. The driving breaks the $Z_2$ parity and thereby split the crossings arising from this symmetry. We have numerically checked this statement, namely that the driving split the crossings between energies with different parities. Furthermore, in fig.~\ref{fig2} we already saw some tendencies of level repulsion. Hence, also according to this definition, the driven RM seems quantum non-integrable.  

\item {\it Solvability.} As we argued above, the question whether the RM is exactly solvable or not is still open. In ref.~\cite{moroz}, the conclusions is that the RM is only quasi-exactly solvable. This means that some properties, but not all, are obtainable analytically. Note that solvability of the RM does not automatically imply solvability of the driven RM.   

\end{enumerate} 

Summarizing, according to the standard definitions of quantum integrability the driven RM should not be considered integrable. This said, it does not mean that the driven RM is not integrable. Of course, as long as there is no accepted definition for quantum integrability we simply do not know if the driven RM is integrable or not. Naturally, the same applies to any model. Notwithstanding, the consensus is to link integrability with quantum thermalization. The absence of a proper definition of integrability makes such a statement ambigouos. The idea of the following section is to underline the obscurity in connecting non-integrability with thermalization.

\subsection{Thermalization of the Driven Rabi Model}
We have seen that our model Hamiltonian should, following the definitions above, be considered non-integrable and, moreover, its semi-classical counterpart is chaotic. Still, as we will show, we find no evidences for thermalization.  

\begin{figure}[h]
\centerline{\includegraphics[width=10cm]{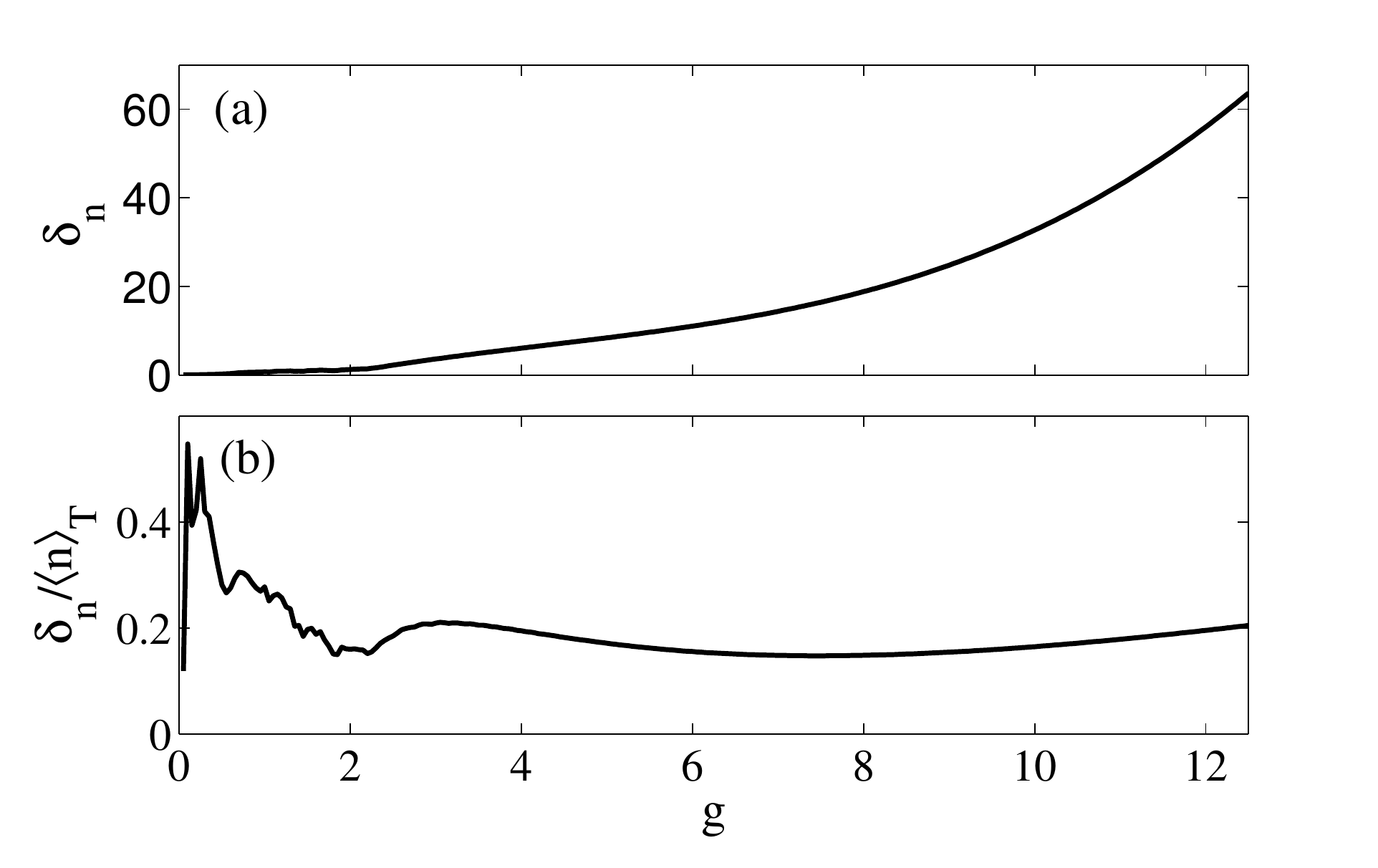}}
\caption{The boson variance $\delta_n$ (a) and the scaled boson variance $\delta_n/\langle n \rangle_T$. The parameters are the same as in fig.~\ref{fig1}. Non-vanishing variance is a manifestation of non-equilibration.} \label{fig3}
\end{figure}

The numerics is carried out using diagonalization of the truncated Hamiltonian. The truncation in the computational basis $\{|n,1\rangle,\,|n,2\rangle\}$ consists in having an upper limit $N_\mathrm{tr}$ of the number of bosons (i.e $n\leq N_\mathrm{tr}$). $N_\mathrm{tr}$ is taken such that our results have converged, i.e. do not depend on $N_\mathrm{tr}$. As local observables we consider $\hat{n}=\hat{a}^\dagger\hat{a}$, $\hat{x}$, $\hat{p}$ and $\hat{\sigma}_\alpha$ ($\alpha=x,\,y,\,z$), and for non-local observables the ``interaction energy'' $\hat{x}\hat{\sigma}_x$. We will only present statistics of the boson number $\bar{n}(t)=\langle\psi(t)|\hat{n}|\psi(t)\rangle$. Similar results are obtained for the other observables. All our simulations of out-of-equilibrium dynamics emerge from a {\it quantum quench}. We prepare the system in the ground state of one Hamiltonian $\hat{H}_0$ and at time $t=0$ we suddenly shift the parameters of the Hamiltonian to a new one $\hat{H}$ under which the state evolves. The ``initial'' Hamiltonian $\hat{H}_0$ is the RM with $g=0.1$ (and thus $\lambda=0$), while the system Hamiltonian $\hat{H}$ of eq.~(\ref{drabi}) typically has $g>1$ in order to be in the highly anharmonic regime and $\lambda\neq0$ in order to break the $Z_2$ symmetry. The initialized state $|\psi(t=0)\rangle$ is predominantly populating eigenstates with zero energy, $\langle\psi(0)|\hat{H}|\psi(0)\rangle\approx0$. In this respect, the eigenstates forming the evolved state are from the irregular part of the spectrum in order to maximize the thermalization effect.    

\begin{figure}[h]
\centerline{\includegraphics[width=10cm]{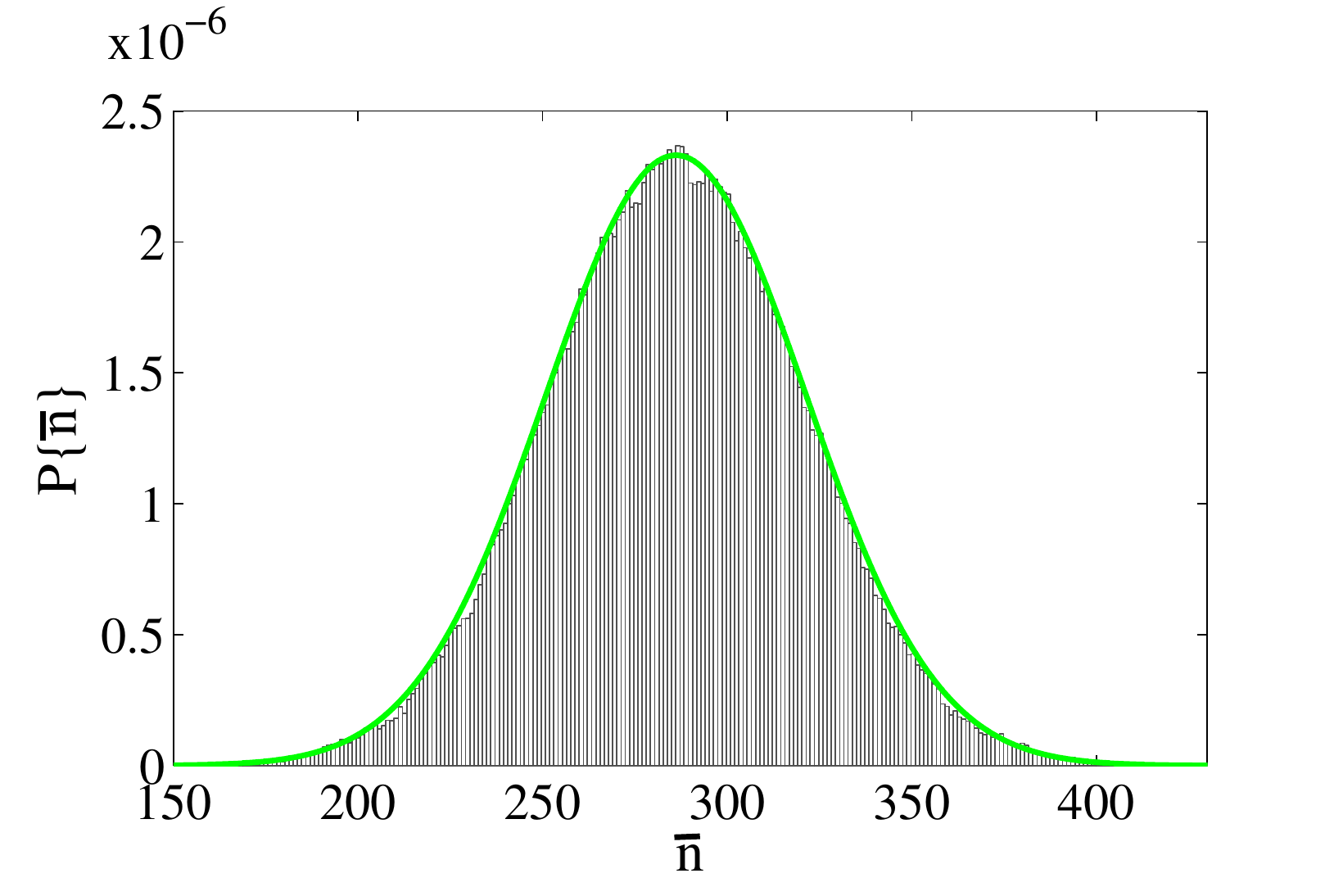}}
\caption{The distribution $P\{\bar{n}\}$ for the driven RM. The solid green line is a a Gaussian curve with mean $\langle n\rangle_\mathrm{T}$ and variance $\delta_n$. The Gaussian shape signals an incommensurability of the eigenvalues $E_\nu$. The parameters are the same as for fig.~\ref{fig1}. } \label{fig4}
\end{figure}

Quantum thermalization implies that the boson variance 
\begin{equation}
\delta_n^2=\lim_{T\rightarrow\infty}\frac{1}{T}\int_0^Tdt\,\bar{n}^2(t)-\left[\lim_{T\rightarrow\infty}\frac{1}{T}\int_0^Tdt\,\bar{n}(t)\right]^2
\end{equation}
should vanish up to order $\mathcal{O}(\eta_\psi)$. How the variance depends on the coupling strength is displayed in fig.~\ref{fig3} (a) for the same parameters as in figs.~\ref{fig1} and \ref{fig2}. For small coupling values $g$ there is some complicated $g$-dependence, while for larger values the variance $\delta_n\sim g^2$. One could imagine that the increased variance for larger $g$'s derives from larger number $\bar{n}(t)$ of bosons. In order to check that this is not the case we show in fig.~\ref{fig3} (b) the scaled variance $\delta_n/\langle n \rangle_\mathrm{T}$ where $\langle n\rangle_\mathrm{T}=\lim_{T\rightarrow\infty}\frac{1}{T}\int_0^Tdt\,\bar{n}(t)$ is the time-averaged boson number. Even the scaled variance does not seem to approach zero but some finite value for large couplings. In a mean-field approach we can understand why the scaled variance goes towards some non-zero value. Within the BOA and deep in the ultrastong coupling regime the ground state of the driven RM will be a coherent state with amplitude $\alpha$ corresponding to the minimum of the lower adiabatic potential $V_\mathrm{ad}^-(x)$~\cite{jonas3,irish2}. For large couplings $g>\omega,\,\lambda$, the coherent amplitude $\alpha=x=g/\sqrt{2}$ so that $\langle n\rangle_T\sim g^2$, and since both $\delta_n$ and $\langle n\rangle_T$ scale as the square of the coupling their ratio should be constant. 

\begin{figure}[h]
\centerline{\includegraphics[width=8cm]{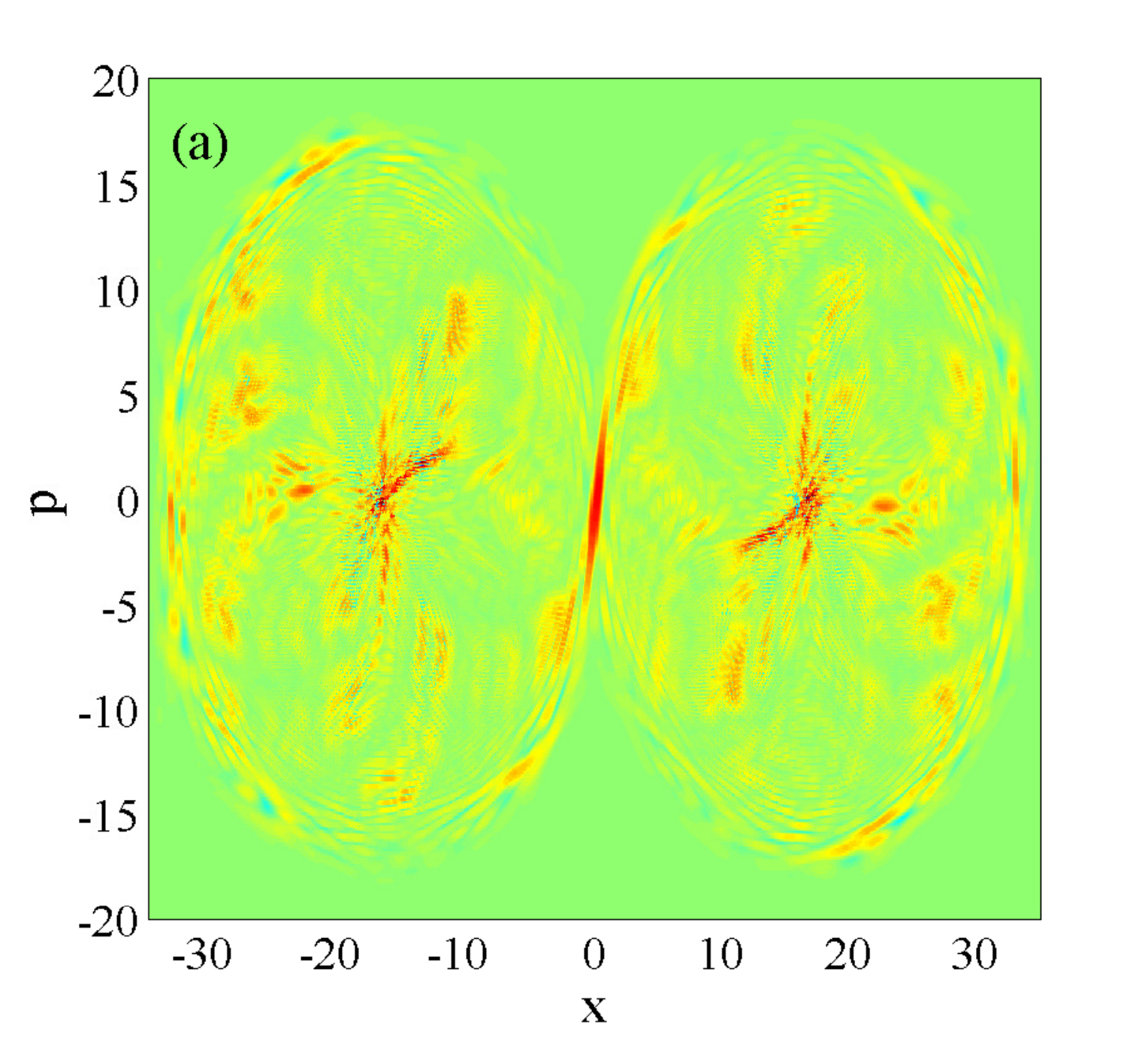}}
\centerline{\includegraphics[width=8cm]{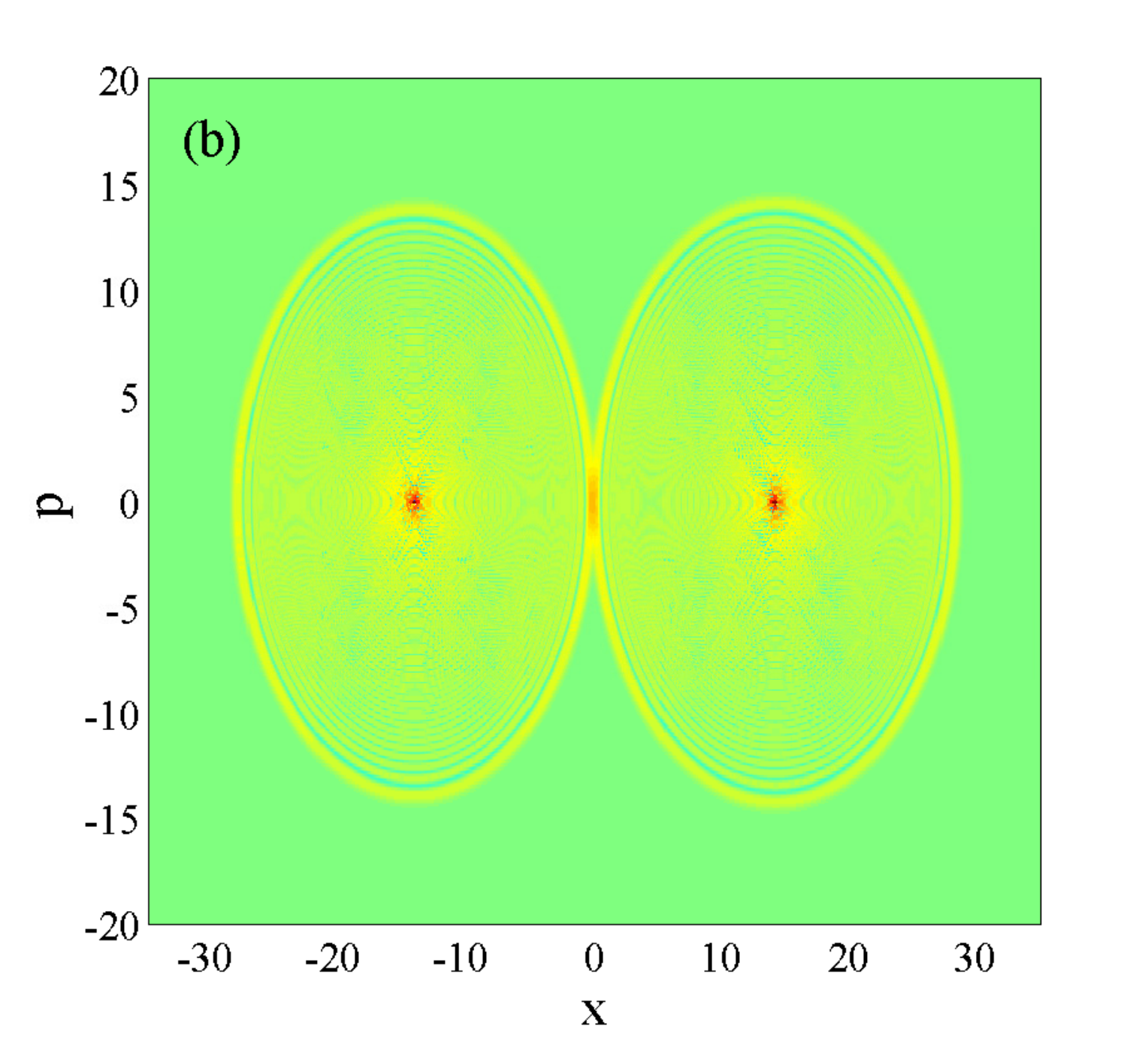}}
\caption{The Wigner distribution $W(x,p)$ of the field state $\hat{\rho}_\mathrm{f}(t)$ for the evolved state after a time $t=500\,000$ (a) and for an eigenstate with eigenenergy $\sim0$ (b). Some of the regular interference structures seen in the eigenstate (b) survives also in the time evolved state of the upper plot. The parameters are the same as for fig.~\ref{fig1}. } \label{fig5}
\end{figure}

We continue analyzing the eigenvalue statistic by recalling a result by Kac~\cite{kac}. Given a set of real values $\{\lambda_\nu\}$ that are incommensurate, that is $\sum_\nu n_\nu\lambda_\nu\neq0$ for any integers $n_\nu$ (except the trivial case $n_\nu=0$ $\forall\,\,\nu$), we form the function $S_\nu(t)=\sqrt{\frac{2}{\nu}}\sum_{j=1}^\nu\cos(\lambda_jt)$. The function $S_\nu(t)$ has a normalized time average $\overline{S_\nu^2(t)}=1$. Letting $\nu\rightarrow\infty$, Kac proved that the probability to find $S_\infty(t)$ between two values $a$ and $b$ is Gaussian, i.e.
\begin{equation}
P\{a\leq S_\infty(t)\leq b\}=\frac{1}{\sqrt{2\pi}}\int_a^bdx\,e^{-x^2/2}.
\end{equation}
From this we expect that for incommensurate eigenvalues $E_\nu$, $\bar{n}(t)$ should be Gaussian. Thus, sampling $\bar{n}(t)$ at random time instants $\{t_\nu\}$ would result in a normal distribution. For the same initial state as in previous figures, we have verified this randomness for the driven RM by calculating the distribution $P\{\bar{n}\}$ as shown in fig.~\ref{fig4}. The fit to a Gaussian with mean $\langle n\rangle_\mathrm{T}$ and variance $\delta_n$ is almost perfect. Interestingly, the Gaussian distribution has also been verified for the Jaynes-Cummings model which is definitely integrable~\cite{stig}. Thus, Gaussianity in this respect does not prove non-integrability nor chaos.

One signature for thermalization is that the evolved state $|\psi(t)\rangle$ is ergodic and shows seemingly irregular phase-space structures~\cite{altland}. For the reduced density operator of the boson field, $\hat{\rho}_\mathrm{f}(t)=\sum_{j=1,2}\langle j|\psi(t)\rangle\langle\psi(t)|j\rangle$, we introduce the Wigner distribution~\cite{mandel}
\begin{equation}
W(x,p,t)=\frac{1}{\pi}\int dy\,\langle x-y/2|\hat{\rho}_\mathrm{f}(t)|x+y/2\rangle e^{ipy}.
\end{equation}
The Wigner distribution is normalized and the marginal distributions agree with the quadrature distributions of the boson field. It is not, however, a proper probability distribution since it is not positive definite. One peculiar property of the Wigner distribution, also demonstrating that it is not a good probability distribution, is that sub-Planck structures are allowed~\cite{zurek}. In fig.~\ref{fig5} (a) we show an example of the evolved Wigner distribution for the same parameters as earlier figures. The time is chosen such that the ``collapse'' of the initially localized distribution has occurred long before the time of the plot. What becomes clear is that the Wigner distribution still shows regular interference structures which is expected for non-chaotic time evolution. We have also calculated the corresponding Wigner distributions for eigenstates of the driven RM for energies around $E\approx0$. A typical example is shown in fig.~\ref{fig5} (b). While the time evolved Wigner distribution is more irregular than the eigenfunction Wigner distribution, some remnants of the symmetric interference structures survive the evolution.

\begin{figure}[h]
\centerline{\includegraphics[width=10cm]{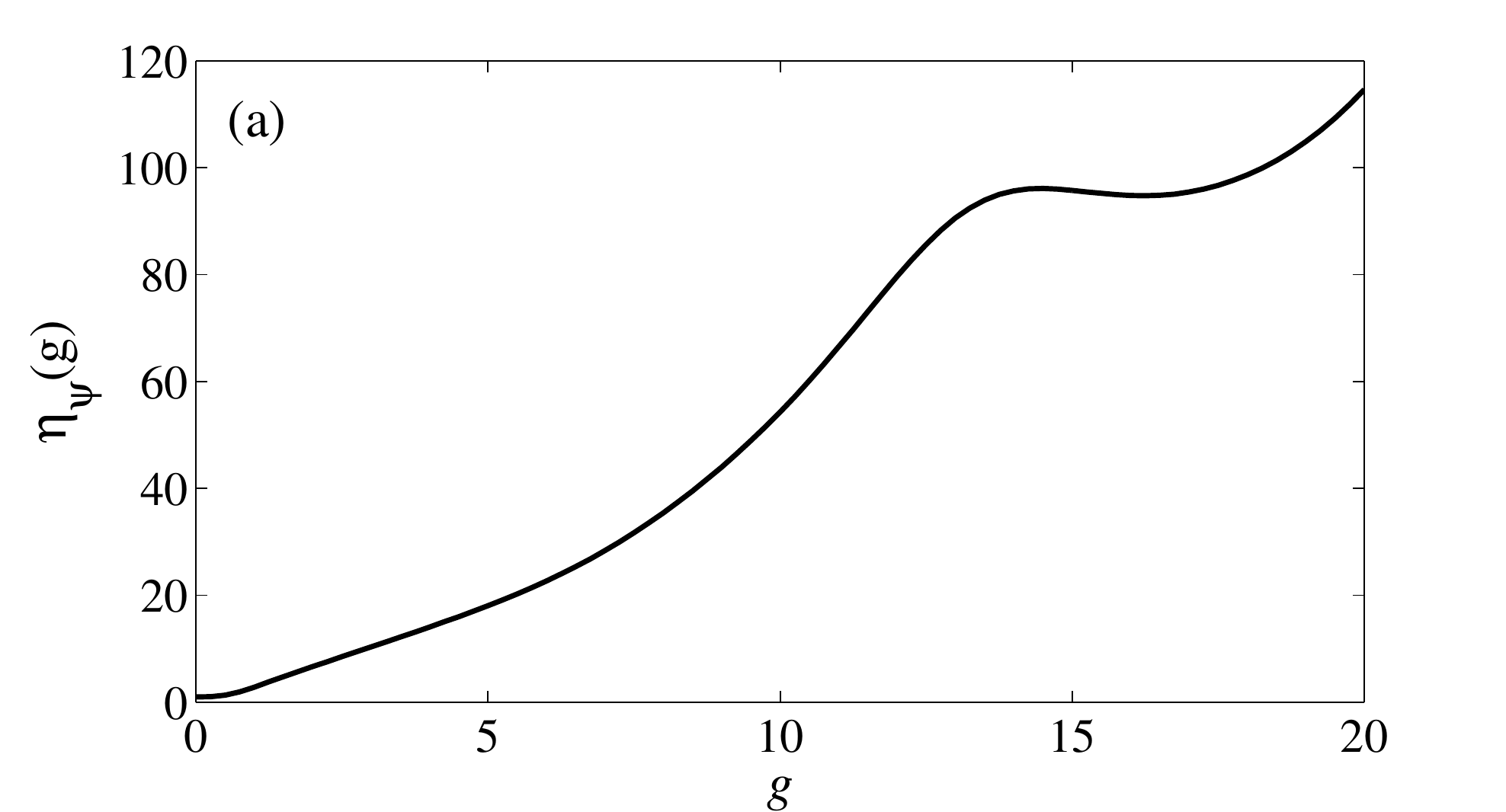}}
\centerline{\includegraphics[width=10cm]{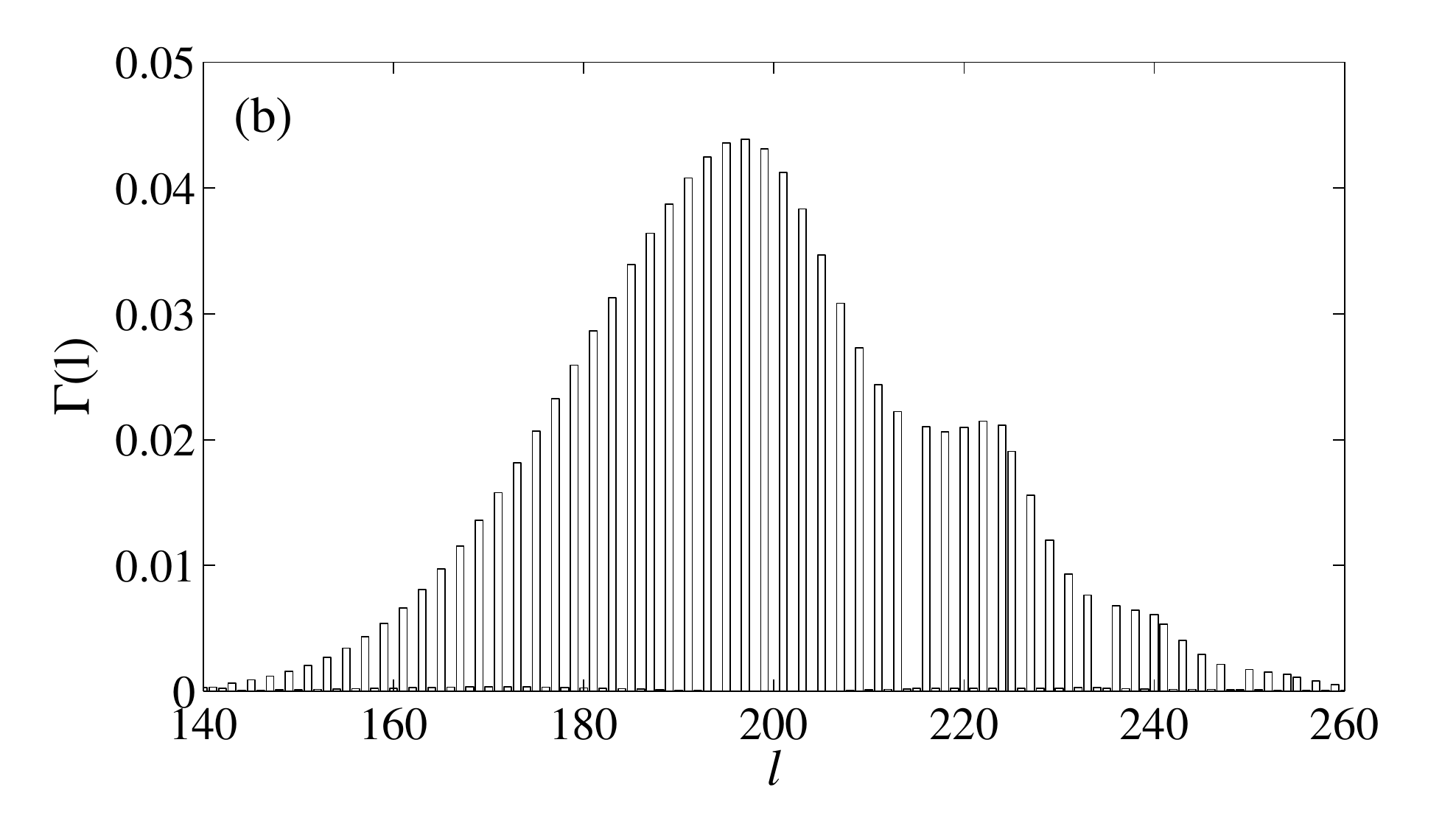}}
\caption{The inverse participation ratio $\eta_\psi$ as a function of the coupling $g$ (a), and the population $\Gamma(l)$ of initial eigenstates $l$ (b). Note that for $g=10$ $\eta_\psi$ estimates $\sim60$ states to be populated which is consistent with (b) calculated for exactly $g=10$. The distribution $\Gamma(l)$ demonstrates that the state is not populating eigenstates at the edge of the spectrum. The unspecified parameters are as in fig.~\ref{fig1}.} \label{fig6}
\end{figure}

All numerical results so far suggest that the driven RM does not show quantum thermalization. However, one may argue that: ($i$) only a specific initial state has been considered, ($ii$) the driven RM is not a many-body model and absence of thermalization could stem from too few contributing states of the sum (\ref{expand}), and ($iii$) if the initial state populates only states at the edges of the spectrum thermalization is not expected~\cite{qcth2}. In order to rule out the first possibility, we have checked for several different initial states. In principle, for a system that thermalizes the expectations $\langle\hat{A}\rangle$ should not depend on details of the initial state but only depend weakly on the system energy $E$. We have thereby focused on analyzing various initial states with different energies. Only states with $E>0$ are interesting since this is were the spectrum is the most irregular. $E$-dependence in $\delta_n$ is indeed found, and in all our numerical simulations we encounter large fluctuations in $\bar{n}(t)$. Thus, we can rule out option ($i$). To get a feeling for the finite size effects of our simulation we calculate the inverse partition ratio~(\ref{partrat}) for different couplings $g$ and the same type of initial quenched states. The results are shown in fig.~\ref{fig6} (a). As expected, $\eta_\psi$ increases for large couplings. If the absence of thermalization derives from finite size effects we should have a decrease of $\delta_n/\langle n\rangle_T$ for increasing $g$ since corrections from zero should scale as $1/\eta_\psi$. This is not what fig.~\ref{fig3} (b) suggests and we thereby cannot explain the large fluctuations in the variance $\delta_n$ as a result of finite size effects. Finally, to check whether our initial state only populates eigenstates at the edge of the spectrum (i.e. for small energies) we plot in fig.~\ref{fig6} (b) the distribution 
\begin{equation}
\Gamma(l)=|\langle\varphi_l|\psi(t)\rangle|^2,
\end{equation}
where $|\varphi_l\rangle$ is the $l$'th eigenstate of the quenched Hamiltonian $\hat{H}$. The distribution is peaked around the 200'th eigenstate, and the first 140 eigenstates are minimally populated. This imply that absence of thermalization is not an outcome of considering an initial state at the edge of the spectrum. Not only states populating the edges of the spectrum can render regular evolution. As for chaotic classical models there might exist ``islands'' in parameter space of regular time evolution also in quantum models~\cite{altland}. However, varying the initial state such situations have not been encountered in this study of the driven RM. This does not prove absence of regular ``islands'' but rather say that if they exist they must be rare.

\section{Concluding remarks}
By considering the driven RM we discussed some ambiguities of quantum integrability and thermalization. Following the most commonly used definitions of quantum integrability, the driven RM would be classified as non-integrable. The fact that there have been claims that the driven RM is solvable~\cite{braak} strengthen the knowledge that quantum integrability is a subtle issue. The solvability of the RM, yet alone the driven RM, has however been questioned~\cite{moroz}. Instead of being exactly solvable, it is argued that only part of the solutions are analytically obtainable, i.e. the model is quasi-exactly solvable. As a non exactly solvable model, a natural exploration is whether the driven RM quantum thermalizes. All our numerical simulations indicated that the model do not thermalize. This, on the other hand, proposes that quantum non-integrability is not a necessity for quantum thermalization. Our findings also hint that classical chaos cannot be taken as a requirement for quantum thermalization.  

It would be interesting to pursue similar analyses for other models that are in some sense quasi solvable. One example would be the Heisenberg $XY\!Z$ spin-1/2 chain. This model only constitute discrete symmetries, but some results, like the ground state energy, can be obtained analytically~\cite{baxter}. Whether this quasi solvability implies lack of quantum thermalization is not known. For the $XY\!Z$ chain including an external field~\cite{fernanda} there exists no known solutions, and thermalization properties of the $XYZ$ model might thereby change in the presence of a field. 
  
\begin{acknowledgments}
The author acknowledges support from the Swedish research council (VR).
\end{acknowledgments}


\begin{thebibliography}{10}
\bibitem{com1} For Hamiltonian systems, as considered in this work, the total energy is naturally conserved.
\bibitem{com2} Note that there exist no rigourous proof that ETH is a general property of quantum thermalizing systems. For example, the eigenstate randominization hypotesis~\cite{erh} is a generalization of ETH.
\bibitem{com4} By neglecting entanglement between the two subsystems, the spin state remains normalized and we can describe it fully with two parameters.
\end{thebibliography}
\end{document}